\documentclass[12pt,preprint]{aastex}

\shorttitle{Altair}
\shortauthors{Buzasi et al.}

\begin{document}


\title{Altair: The Brightest Delta Scuti Star}


\author{D. L. Buzasi\altaffilmark{1}, H. Bruntt\altaffilmark{1,2}}
\email{derek.buzasi@usafa.af.mil,bruntt@phys.au.dk}
\author{T. R. Bedding\altaffilmark{3}, A. Retter\altaffilmark{3,4}}
\email{bedding@physics.usyd.edu.au, retter@astro.psu.edu}
\author{H. Kjeldsen\altaffilmark{2}}
\email{hans@phys.au.dk}
\author{H. L. Preston\altaffilmark{1}, W. J. Mandeville\altaffilmark{1}}
\email{heather.preston@usafa.af.mil,jody.mandeville@usafa.af.mil}
\author{J. Catanzarite\altaffilmark{5}, T. Conrow\altaffilmark{6}, 
R. Laher\altaffilmark{7}}
\email{Joseph.H.Catanzarite@jpl.nasa.gov, tim@ipac.caltech.edu, laher@ipac.caltech.edu}

\altaffiltext{1}{Department of Physics, US Air Force Academy, 2354 Fairchild
Dr., Ste. 2A31, USAF Academy, CO}
\altaffiltext{2}{Theoretical Astrophysics Center, University of Aarhus,
8000 Aarhus C, Denmark}
\altaffiltext{3}{School of Physics, University of Sydney, Sydney, NSW 2006,
Australia}
\altaffiltext{4}{Department of Astronomy and Astrophysics, Pennsylvania State University, 
525 Davey Laboratory, University Park, PA 16802}
\altaffiltext{5}{Interferometry Science Center, California
Institute of Technology, Pasadena, CA 91125}
\altaffiltext{6}{Infrared Processing and Analysis Center, MS 100-22, California
Institute of Technology, Pasadena, CA 91125}
\altaffiltext{7}{Spitzer Science Center, MS 314-6, California
Institute of Technology, Pasadena, CA 91125}


\begin{abstract}
We present an analysis of observations of the bright star Altair ($\alpha$ Aql)
obtained using the star camera on the Wide-Field Infrared Explorer (WIRE) satellite.
Although Altair lies within the $\delta$ Scuti instability strip, previous
observations have not revealed the presence of oscillations. However, the WIRE
observations show Altair to be a low-amplitude ($\Delta m < 1$~ppt) $\delta$ Scuti star
with at least 7 modes present.
\end{abstract}




\section{Introduction}
Delta Scuti stars are a variety of pulsating variable star located within the classical 
instability
strip of the HR diagram (Rodriguez \& Breger 2001). 
They inhabit the region between $3.8 < \log \rm T_{eff} < 3.95$ and
$0.6 < \log \rm (L/L_\sun) < 2.0$, and display periods ranging from 0.02 to 0.3 days.
Some belong to the high-amplitude $\delta$ Scuti (HADS) class, which display $V$ amplitudes
in excess of 0.3 mag, and generally oscillate in radial modes, while the lower-amplitude
members of the class have more complex frequency structure, typically showing numerous
nonradial modes.

Altair ($\alpha$ Aql) is an A7 IV-V main sequence star (Johnson \& Morgan 1953)
and is the 12th brightest star in the sky (V = 0.755, Cousins 1984). While it lies inside
the instability strip, we are unaware of any reports of photometric variability.
Erspamer \& North (2003) report $\rm T_{eff} = 7550 K$ and $\log g = 4.13$, and
Zakhozhaj (1979, see also Zakhozhaj \& Shaparenko 1996) derives a mass of $1.75 M_\sun$ 
and a radius of $1.58 R_\sun$ based on photometry.
The Hipparcos parallax for Altair is $194.44 \pm 0.94 \rm~ mas$, giving a distance
of $5.143 \pm 0.025 \rm~pc$. Combined with the observed V magnitude and the bolometric
correction from Flower (1996) gives $\rm M_{bol} = 2.18$. 
Altair is a rapid rotator, with spectroscopically derived 
$v \sin i$ values in the literature ranging from
$190 \rm~km~s^{-1}$ to $250 \rm~km~s^{-1}$ (Royer et al. 2002).

Altair is nearby enough to make direct measurement of its diameter possible. Richichi et al.
(2002) reported a diameter of $3.12 \rm ~mas$, corresponding to a radius of $1.72 \rm R_\sun$.
Recent interferometric observations
(Van Belle et al. 2001) have established that Altair is oblate, with equatorial diameter
of $3.46 \rm ~mas$ (corresponding to a radius of $1.9 \rm R_\sun$) 
and polar diameter of $3.037 \rm ~mas$, for an axial ratio
$a/b = 1.14 \pm 0.029$.
Van Belle et al. also derive a value for $v \sin i$ of $210 \pm 13 \rm~
km~s^{-1}$.

Altair's absolute V magnitude is +2.22 (based on the Hipparcos distance) and
its $\rm (b - y)_0 = 0.137$
(Hauck \& Mermilliod 1998; Nekkel et al. 1980), which places it
well inside the $\delta$ Scuti instability strip (Breger 1990). However, like many
such objects, Altair has never been detected to oscillate. It is as yet unclear whether
these non-oscillating stars within the instability strip are truly non-oscillating,
or whether the non-detections are instead due to either periods or amplitudes which 
are undetectable from the ground; some indications are that slow rotation is a necessary
condition for large oscillation amplitudes (Breger 1982, Solano \& Fernley 1997,
 but for an alternate view
see Rasmussen et al. 2002).

\section{Observations and Data Reduction}
The Wide-Field Infrared Explorer (WIRE) mission was launched by NASA into a
sun-synchronous orbit on 4 March 1999. The primary mission failed several days
after launch, but conversion of the spacecraft into an asteroseismology platform
using the onboard star camera as the new science instrument began on 30 April 1999.

The star camera consists of a 52 mm, f/1.7 refractive optic feeding a $512^2$ SITe
CCD and a 16-bit analog-to-digital converter. The camera has a field of view of roughly
7.8 degrees square, corresponding to an image scale of 1 arc minute per pixel; it
is unfiltered, but the effective response is roughly $\rm ( V + R )$. Observations of
Altair were carried out with a cadence of 0.5 seconds. Details
of the instrument and its performance can be found in Buzasi (2000, 2002, 2003). 

Altair was observed from 18 October through 12 November 1999. Exposures of 0.1 second were taken
during about 40\% of the spacecraft orbital period. Some additional gaps due to mission
operations constraints were also present, resulting in an overall duty cycle of approximately
27\%. 

We have reduced the data using two independent techniques, which we now describe.
In the first approach, as described in Buzasi (2000), data reduction began by applying a simple 
aperture photometry algorithm, which simply involved summing the central 
$4 \times 4$ pixel region of the $8 \times 8$ pixel field of view. The background 
level, due primarily to
scattered light from the bright Earth, was estimated based on the four corner
pixels of each image, and subtracted. 
A total of 
approximately 1.27 million data points were acquired during the observing window. 

Following aperture photometry, a clipping algorithm was applied to the 
time series, resulting in the rejection of any points deviating more than 
$2.5\sigma$ from the mean flux, image centroid, or background level. 
In practice, at this point in the data reduction process, the rms noise in the 
signal is extremely large 
($\sigma \approx 12$~ppt) 
due to incomplete subtraction of scattered light, so clipping 
in this way only removes truly deviant points. The majority of these are due to 
either data from the wrong 8x8 CCD ``window'' being returned, or to data acquired 
before the spacecraft pointing had settled during a particular orbital segment.
After this stage, approximately 1.21 million data points remained.

As noted in the previous paragraph, the crude scattered light removal 
discussed above is clearly inadequate. 
Much of the difficulty inherent
in the reduction of WIRE photometric data is in successfully characterizing and 
removing the scattered light signature (Buzasi 2000). We phased 
the data at the satellite orbital period, binned the result in steps of 
phase $10^{-3}$, and fit and subtracted a spline to the result, which lowered the
rms noise level by approximately an order of magnitude. For an example of this
process, see Retter et al. (2003).

The resulting distribution of magnitudes is a Gaussian with some outliers. We discarded
466 points that were more than $4 \sigma$ from the mean, and then binned into $50 \rm~s$
intervals, discarding bins containing fewer than 20 points. To remove slow trends, we
also fitted and subtracted a third-order polynomial. The final binned time series contained 
9828 points. 

In the second approach we first fit a 2D-Gaussian profile
to each $8 \times 8$ image in order to examine
changes in the central $(x,y)$ position of the stellar
profile and to find out how the FWHM changes.
The $(x,y)$ position of the main target is found to be
extremely stable, ie. $x = 3.271\pm0.006$ and $y = 3.158\pm0.008$,
where $(x,y) = (3.5, 3.5)$ is the center of of the $8 \times 8$ window.
The FWHM of the stellar profile is also very stable, $F = 1.860\pm0.003$ pixels.

For each image we measured the stellar flux using 
aperture photometry. The sky level was determined by 
the average flux of the four pixels in the corners of 
the CCD image, and the aperture was centered on the 
$(x,y)$ position found from the Gaussian 2D-fit. 
The aperture sizes were scaled with the FWHM of
the fitted Gaussian to ensure that approximately 
the same region of the PSF profile is sampled.
To ensure robustness, the sky background, the $(x,y)$ position,
and the FWHM we used were the average of the 
preceding 20 and following 20
images, thus averaging over a full 20 seconds. 

We used eight different apertures with increasing radii, and
found the lowest noise when applying an
aperture with radius $r=1.39 F$, corresponding 
to $r = 2.52$ pixels.
 
We binned the light curve into groups of $n=51$ data points, requiring
that each $(x,y)$ position be within 0.1 pixels of the mean;
3\,$\sigma$ outliers were removed. 
We also required that the time interval between
subsequent data points not exceed 10 seconds.
The resulting light curve has 40\,088 data points. 
The original data points have a point to point noise 
of 0.73~ppt, while in the binned light curve the noise is 0.16~ppt.


We note that after subtracting the detected oscillation modes (cf. Section 3)
we detected a slight correlation with the FWHM and have 
removed this by fitting a second degree polynomial to
the residual flux vs.\ FWHM. 

\begin{figure*}
  \begin{center}
        \includegraphics[width=17cm]{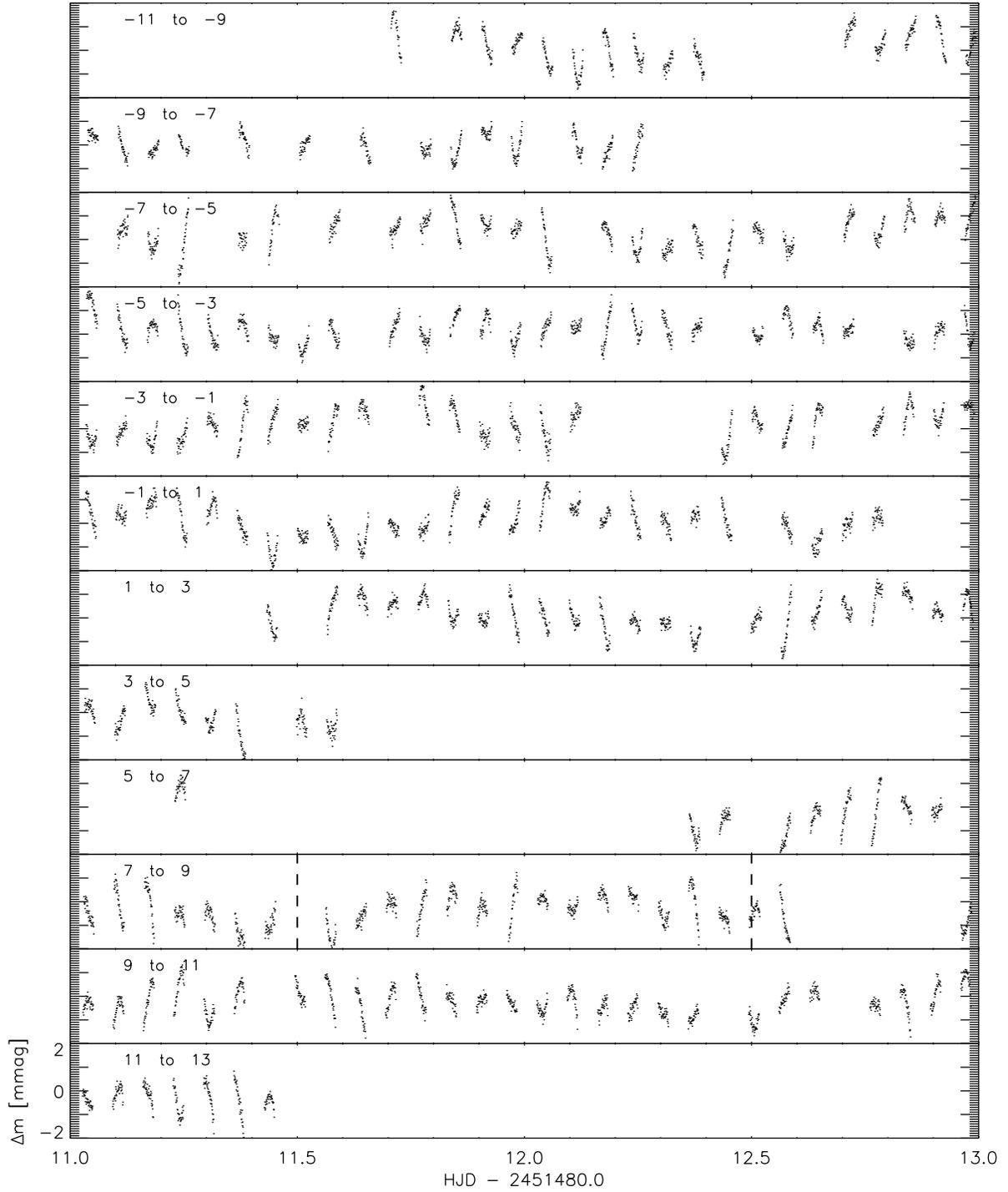}
        \caption{The complete observed light curve of Altair. 
The time is the
heliocentric Julian date relative to the time $t_0 = 2\,451\,480$,
and individual panels are labeled according to range of days they
encompass relative to that date.
        \label{fig:lc_full}}
   \end{center}
\end{figure*}

\begin{figure*}
  \begin{center}
        \includegraphics[width=12.5cm]{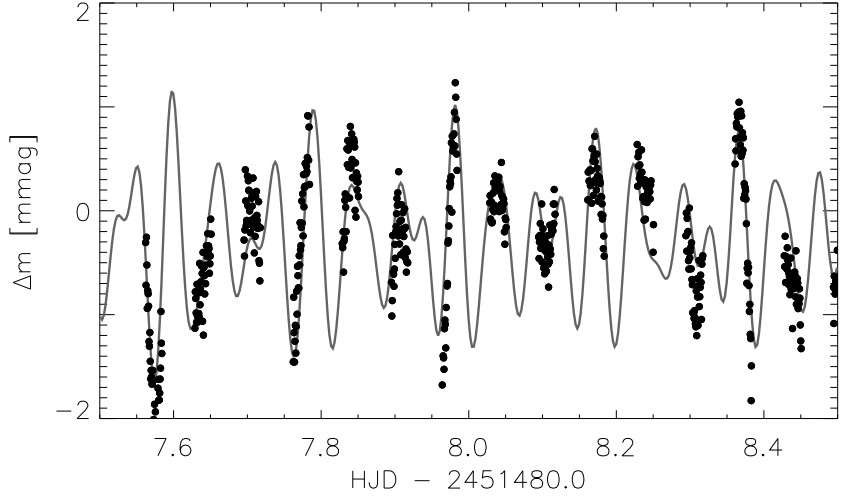}
        \caption{18 hours of the observed light curve of Altair. 
The solid curve is the fit to the light curve, based on the frequencies shown
in Table 1. The time is the
heliocentric Julian date relative to the time $t_0 = 2\,451\,480$.
        \label{fig:lc}}
   \end{center}
\end{figure*}

The entire Altair light curve is shown in Figure~\ref{fig:lc_full}, while
Figure~\ref{fig:lc} zooms in on a portion. In the latter figure,
the fit is overlaid as a solid curve.

\begin{figure*}
        \includegraphics[width=8.5cm]{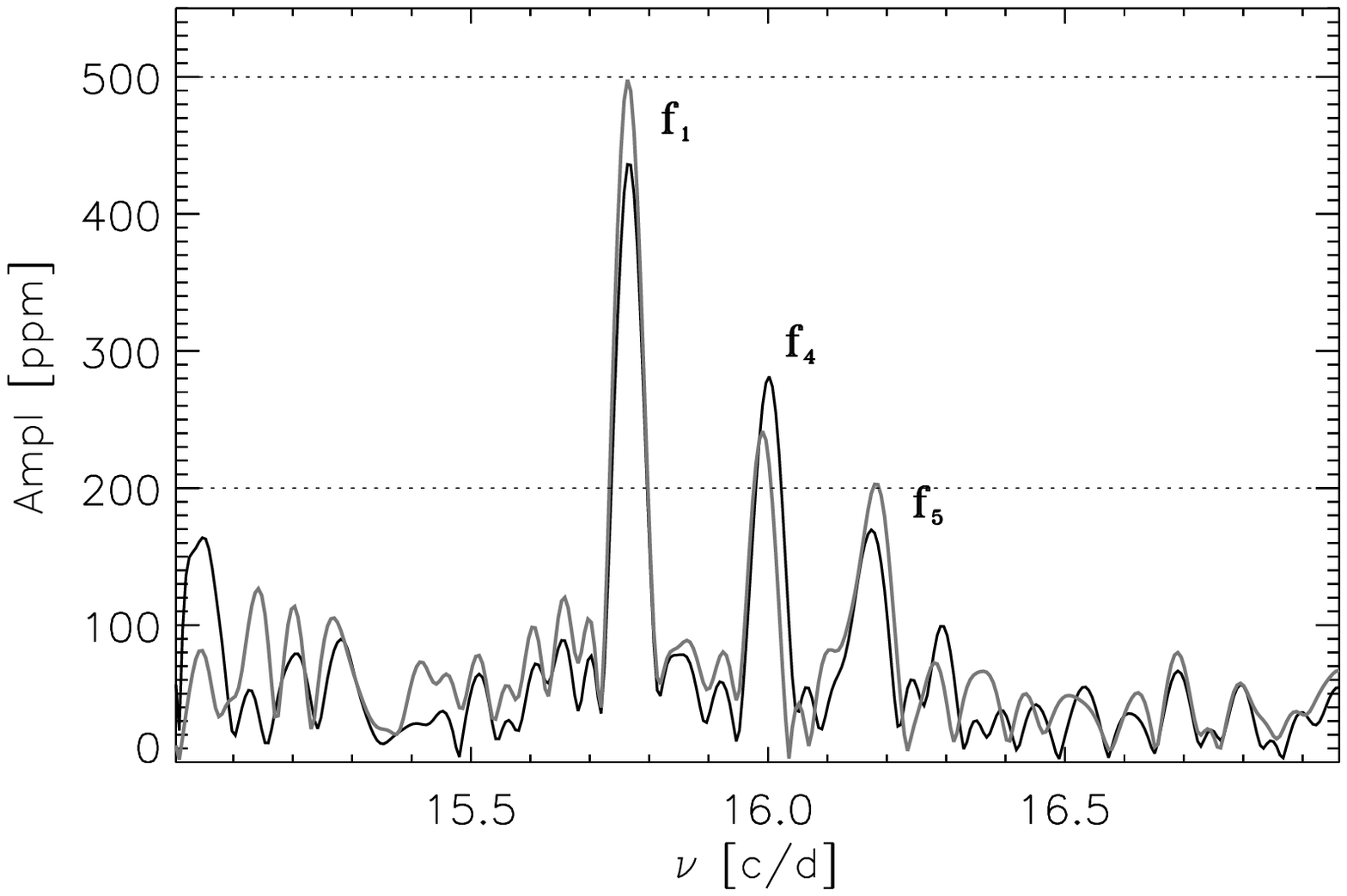}
        \includegraphics[width=8.5cm]{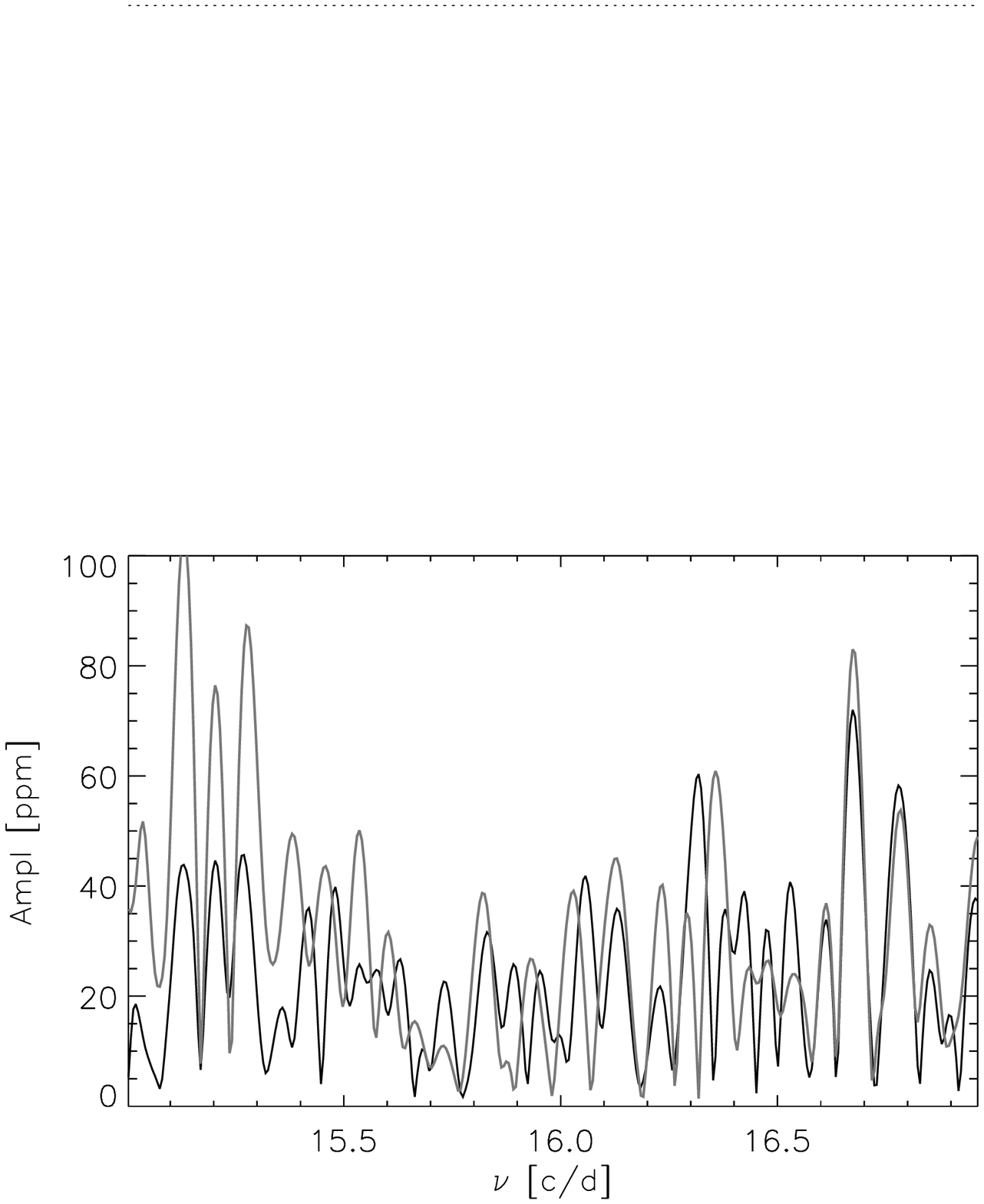}
        \caption{The {\em left} panel shows the amplitude spectra of
the light curve of Altair in the range 15.3--16.6~c/day. The gray and
black colors correspond to the first and second data reduction techniques (see text
for details). In the {\em right} panel the three main modes have
been subtracted from the light curve. While the apparent amplitude is
larger for the ``gray'' spectrum the noise level is also significantly
higher. The horizontal dotted lines are added to aid the eye.
        \label{fig:amp}}
\end{figure*}

\section{Analysis}

Figure~\ref{fig:amp}
shows the a portion of the amplitude spectrum for 
the light curve extracted as described in Section 2.
In the left panel the oscillation signal 
around 16.0~c/day can be seen.
The right panel shows the amplitude spectrum when the
three detected modes are removed. 

Note that in both data sets a slow drift has been observed corresponding
to a peak in the amplitude spectrum at a period of 27.0 days
and an amplitude of just 320~ppt. 
Due to aliasing this will give rise to
a group of peaks around $f_{\rm alias} = f_{\rm orb} - f_0$, where
$f_{\rm orb}$ is the orbital frequency of WIRE, ie. 15.00 c/day. 
This is just below the frequency of the 
three observed peaks in the amplitude spectrum and 
with comparable amplitudes.

After removing the oscillation signal from the light curves
we measured the noise level in the amplitude spectrum.
In the high frequency range (30--60~c/day) the Buzasi (2000) approach
(technique 1; gray curve) gives a noise level of 10.5 ppm, while our new approach
(technique 2; black curve) gives a noise level of 9.8 ppm.
In the range 15--17~c/day (around the oscillation peaks) we
find 26.6 and 21.0 ppm, respectively. Thus, the noise levels
are comparable at high frequencies but significantly lower 
for the second method at the lower frequencies. 
The apparently higher amplitude of the three oscillation modes
(see left panel of Figure~\ref{fig:amp})
derived from technique 1 is due to this higher noise level.

\begin{figure*}
        \includegraphics[width=15.0cm]{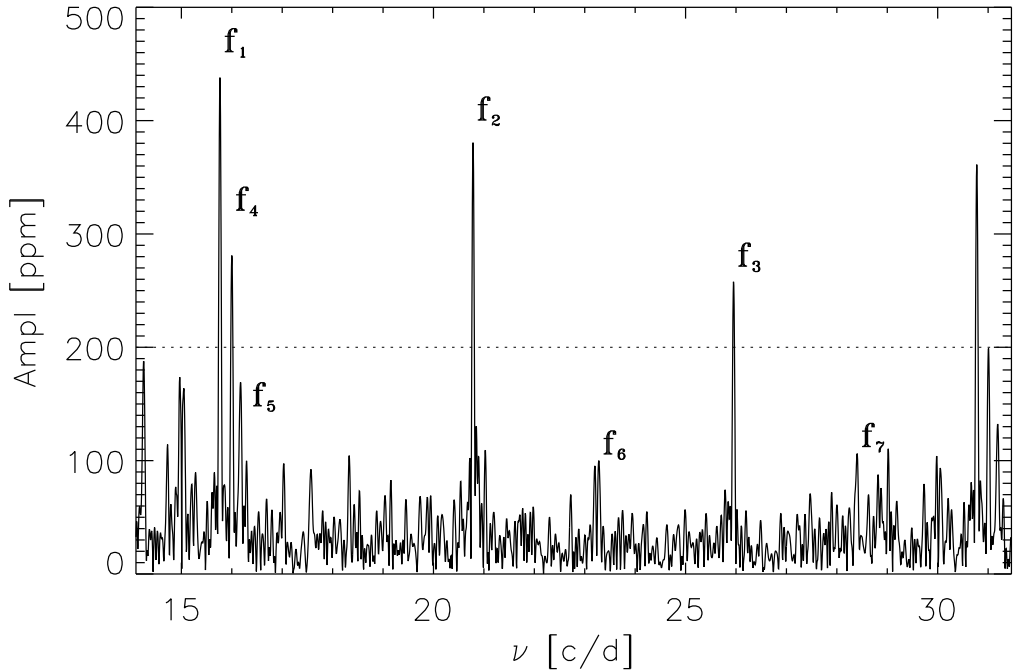}
        \caption{Amplitude spectrum of Altair in the range 
                 15.5 -- 26.5~c/day. The seven modes have been labeled. Note that
the $f_7$ peak lies to the left of the label; compare to Figure~\ref{fig:amp_white}.
 The horizontal dotted line is added to aid the eye.
        \label{fig:amp2}}
\end{figure*}

Figure~\ref{fig:amp2} shows the amplitude spectrum of 
Altair from the second reduction method. We identify
seven modes in this spectrum. Note that both reduction methods yield 
identical oscillation frequencies within the estimated uncertainties.

\begin{figure*}
        \includegraphics[width=15.0cm]{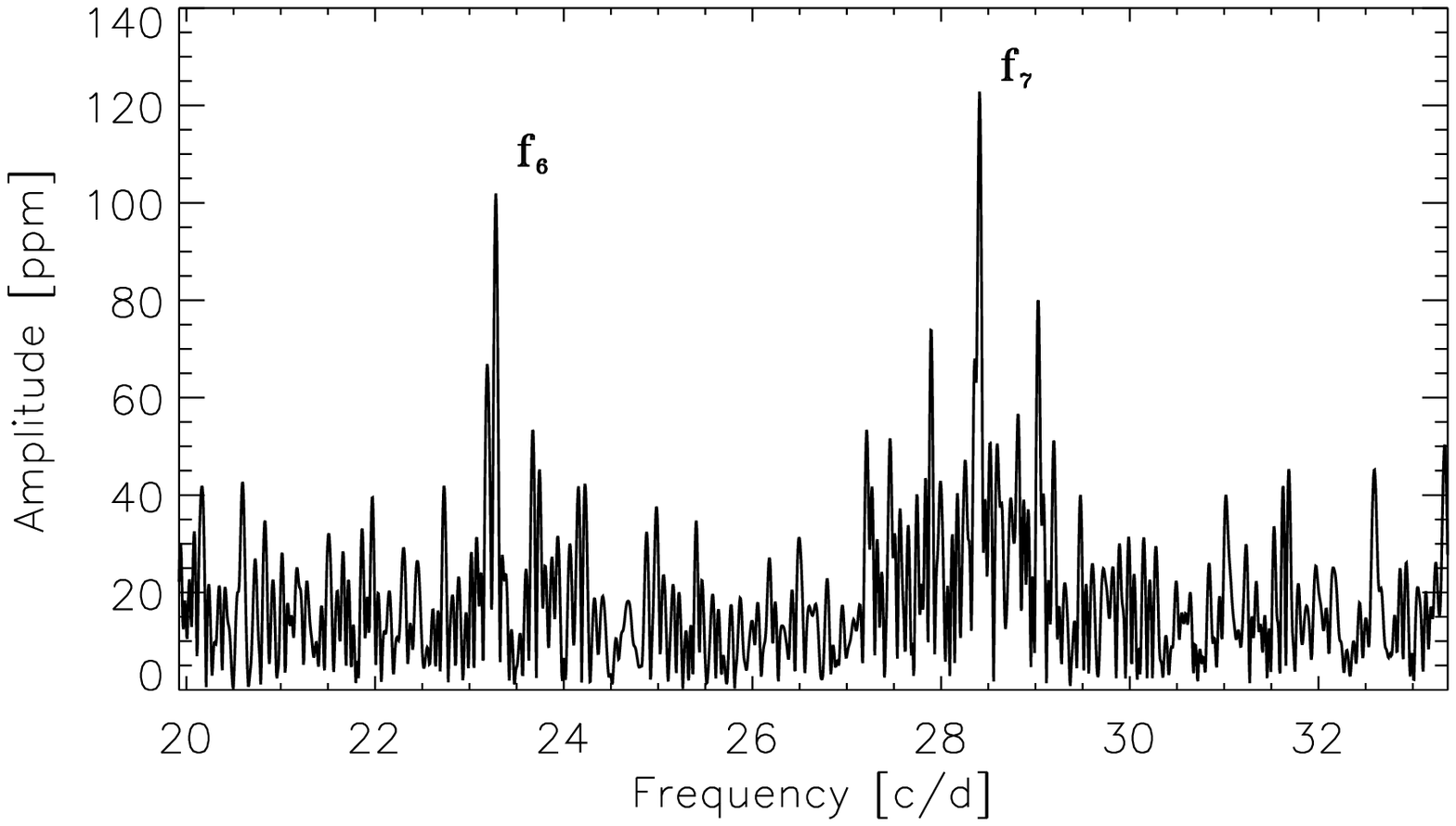}
        \caption{The amplitude spectrum of Altair in the range 
                 15.5 -- 26.5~c/day, prewhitened to remove frequencies $f_1$ -- $f_5$.
The significance of the peaks we identify as $f_6$ and $f_7$ is clearly apparent.	
        \label{fig:amp_white}}
\end{figure*}

Table~\ref{tab:sn} shows for each 
oscillation mode the frequency, amplitude, phase, and signal to noise
ratio for the two different techniques.
We used the software {\tt period98} (Sperl 1998) to determine 
the mode parameters.
The phase corresponds to the heliocentric Julian date
$t_0 = 2\,451\,480$ and the phase $\phi$ is in units of $ 2 \pi$,
thus, the light curve fit shown in Figure~\ref{fig:lc} is 
represented by
$\delta m = \sum_{i=1}^7 A_i * 
\sin (2 \pi [f_i * (t - t_0) + \phi_i ]  ) $.
We estimate amplitude errors to be of the order of $\pm 20$~ppm; this 
is apparent from a comparison of the amplitude of $f_2$ in Figure~\ref{fig:amp2}
to that derived from {\tt period98} and shown in Table~\ref{tab:sn}.
Frequency errors are approximately $\pm 0.001 \rm~c/d$.

\begin{table}
\centering
\caption{Parameters of the observed oscillation modes in Altair.
\label{tab:sn}}
\vskip 0.2cm
\begin{tabular}{l|rrr|rr}
 & $\nu$ [c/d]  & $A$ [ppm] &  $\phi$ &(S/N)$_1$& (S/N)$_2$ \\
\hline
$f_1$  &    15.768 &       413 &   0.383 &    19.7 &    23.3 \\
$f_2$  &    20.785 &       373 &   0.299 &    28.6 &    28.5 \\
$f_3$  &    25.952 &       244 &   0.141 &    19.2 &    18.7 \\
$f_4$  &    15.990 &       225 &   0.689 &     8.7 &    12.6 \\
$f_5$  &    16.182 &       139 &   0.461 &     7.5 &     7.8 \\
$f_6$  &    23.279 &       111 &   0.102 &     8.0 &     8.5 \\
$f_7$  &    28.408 &       132 &   0.553 &     7.4 &     7.4 \\
\end{tabular}
\end{table}


\section{Discussion}

The accepted definition of a $\delta$~Scuti star (Breger 1979, Rodr\'{\i}guez et al. 1994) is 
a variable star with spectral type between A2 and F0, oscillation periods less than 
$0.3 \rm~d$, and visual amplitudes typically less than $0.3 \rm~mag$. SX~Phe stars are
similar to $\delta$~Scuti in most respects, but show Population II characteristics
combined with anomalously large masses and young ages, while $\lambda$ Bootis stars
combine $\delta$~Scuti-type oscillations with abundance anomalies in Fe, C, N, O, and S. 
Since Altair shows neither abundance anomalies
nor Pop II characteristics (indeed,
it lies towards the low-mass end of the instability strip), and has spectral type
A7~IV-V and primary oscillation period 0.0634 days, we identify it as a $\delta$~Scuti star. In the following we will compare the observed oscillation modes with 
a number of published models with parameters close to Altair.  However, we stress that these models do not take into account rotation, which will significantly change the structure of the star. In a future paper we will address the detailed modeling of Altair.

Using the basic pulsation relation
\begin{equation}
P \sqrt{\rho/\rho_\sun} = Q
\end{equation}
where $P$ is the pulsation period, $\rho$ the mean stellar density, and $Q$ the pulsation 
constant, it is straightforward to derive a theoretical period-luminosity relation of the
form (Breger \& Bregman 1975)
\begin{equation}
\log Q = -6.454 - \log f + 0.5 \log g + 0.1M_{bol} + \log T_{e}
\end{equation}
where frequency $f$ is in $\rm d^{-1}$. The
pulsation constant $Q$ varies depending on both stellar structural considerations and
the specific mode in question, but for the radial fundamental mode $Q = 0.033 \rm~d$ for
$\delta$~Scuti stars (Breger 1979). Using our fundamental parameters for Altair from Section 1, 
we then predict the frequency of the fundamental
mode to be $15.433 \rm~d^{-1}$, 
remarkably close to our observed $f_1 = 15.768 \rm~d^{-1}$. 

Extensive linear and nonlinear analysis of $\delta$ Scuti oscillation properties has been
published (see, {\it e.g.} Bono et al. 1997, Milligan \& Carson 1992, Andreasen, Hejlesen,
\& Petersen 1983, Stellingwerf 1979, and references therein). Use of these model
grids allows us to further clarify the mode identification in Altair. 

Stellingwerf (1979) calculated a grid of models intended to represent both dwarf Cepheids
and $\delta$ Scuti stars. His models 4.2 and 4.3 appear to best represent Altair on the basis
of global physical properties. Model 4.2 has $\rm M = 2.0 M_\sun$, $\rm T_e = 7700 \rm K$,
$\rm R = 2.03 R_\sun$, and $\log \rm g = 4.12$, while model 4.3 is somewhat smaller 
($\rm R = 1.88 R_\sun$) and hotter ($\rm T_e = 8000 \rm K$). Both models have $X = 0.7$,
$Z = 0.02$.
Table 2
shows the comparison between our observations and Stellingwerf's models for our
$f_1$, $f_2$, and $f_3$. Overall agreement is quite good, with our observations generally
lying between the model predictions. Stellingwerf's models also predicted that the
third overtone should also be visible, at a frequency between $27.701$ and $31.056 \rm~d^{-1}$.
There is no conclusive evidence in our amplitude spectrum for such a mode, though the location
of our $f_7$ is
certainly suggestive.

Milligan \& 
Carson (1992) calculated a grid of $\delta$~Scuti oscillation frequencies based on evolved
Kippenhahn et al. (1967) interiors models. They label models both by mass and age.
Their $1.6 \rm M_\sun$ models are the best fit to Altair in
terms
of global observable parameters; model m16e02 has
$\rm T_{e} = 7765 \rm K$ and $\log \rm g = 4.21$,
model m16e03 has
$\rm T_{e} = 7560 \rm K$ and $\log \rm g = 4.15$,
and model m16e04 has
$\rm T_{e} = 7285 \rm K$ and $\log \rm g = 4.08$.
The usage is such that m16e01 has mass $\rm 1.6 M_\sun$ and $\log \tau = 2.6$, where age 
$\tau$
is measured in units of $10^6$ years; m16e03 has $\log \rm~age = 2.7$, and m16e04
has $\log \rm~age = 2.8$.
Table~2 gives the comparison between our measured frequencies and those predicted by
Milligan \& Carson for the three largest amplitude modes.

\begin{table*}[h]
  \begin{center}
  \caption{Parameter Comparison to the models of Stellingwerf (1979) and 
Milligan \& Carson (1992). Model 
labels are described in the text.} 
\vskip 0.2cm
    \begin{tabular}[h]{ccccccc}
    \tableline \tableline
Parameter & Measured & 4.2 & 4.3 & m16e02 & m16e03 & m16e04 \\ \tableline
$f_1$ & 15.768 & 14.641 & 16.340 & 18.5235 & 16.6514 & 14.6556 \\
$f_2$ & 20.785 & 18.692 & 20.833 & 24.4240 & 21.8936 & 19.2231 \\
$f_3$ & 25.952 & 23.095 & 25.840 & 30.4406 & 27.2681 & 23.9363 \\
$\Pi_1 / \Pi_0$ & 0.75862 & 0.78328 & 0.78433 & 0.75841 & 0.76056 & 0.76240 \\
$\Pi_2 / \Pi_0$ & 0.60758 & 0.63395 & 0.63235 & 0.60852 & 0.61066 & 0.61227 \\ 
$Q_0$ & 0.03230 & 0.03479 & 0.03117 & 0.03244 & 0.03253 & 0.03256 \\
$Q_1$ & 0.02451 & 0.02725 & 0.02445 & 0.02460 & 0.02474 & 0.02482 \\
$Q_2$ & 0.01963 & 0.02205 & 0.01971 & 0.01974 & 0.01986 & 0.01994 \\ 
\tableline
      \end{tabular}
  \end{center}
\end{table*}

Milligan \& Carson also used nonlinear models to investigate oscillation amplitudes
for a number of their models, including m16e03. In each case, they searched for an
initial velocity amplitude that would give a settled pulsation with constant maximum
kinetic energy per period. For the case of m16e03, they found that the resulting second
overtone
frequency was $26.67 \rm~d^{-1}$ -- similar to that of the linear calculation -- and
that the
required velocity amplitude was $100 \rm~m~s^{-1}$, 
corresponding to an approximate $V$ amplitude of $0.5$~ppt. Both results are
quite close to those determined from our observations. We note that spectroscopic 
detection of such a small velocity amplitude would be quite challenging, though
certainly within the current state of the art (Butler 1998).

Figures~\ref{fig:amp2} and \ref{fig:amp_white}
and Table~1 show the presence of four additional peaks ($f_4$ -- $f_7$) in the amplitude
spectrum with semi-amplitudes greater than $100$~ppm. $f_4$ and $f_5$ are located near the
frequency of the radial fundamental mode, but are well-separated in frequency space.
Given their frequencies, these peaks must correspond to nonradial oscillation modes,
but beyond that our identification is uncertain. The identifications of $f_6$ and $f_7$ are 
likewise uncertain, although we tentatively identify $f_7$ with the third radial overtone.

More modern models of $\delta$ Scuti stars, incorporating OPAL and OP opacities, were
explored by Petersen \& Christensen Dalsgaard (1996, 1999). Updated versions of the
stellar model sequences used in these papers were published online by Petersen (1999),
and we compared our frequencies to these results. Unfortunately, the metallicity of
Altair appears to be somewhat uncertain; the value given in Cayrel de Stroebel et al. (1992) is
actually that of $\alpha$ Aqr rather than $\alpha$ Aql. However, Zakhozhaj \& Shaparenko (1996) and
Erspamer \& North (2003) give $\rm [Fe/H]=-0.24$, while use of the TempLogG 
tool (http://ams.astro.univie.ac.at/templogg/) together with Str\"omgren photometry
gives $\rm [Fe/H]=-0.15$,
so we have chosen to compare our data to Petersen's models with $Z = 0.02$ (equivalent
to $\rm [Fe/H]=-0.16$). 

The effective temperature and bolometric magnitude of Altair fall slightly below Petersen's
$1.8 M_\sun$ evolutionary track, a result which is in accord with Erspamer \& North's value
of $1.75 M_\sun$. In addition, Petersen's Figure 4 allows us to estimate the stellar mass
based on $(\log T_{eff}, \log \Pi_0)$ -- the result once again is $M \approx 1.75 M_\sun$.

Our first harmonic, however, does not fit well onto Petersen's $(\log \Pi_0, \Pi_1 / \Pi_0)$
diagram. In particular, for our value of $\log \Pi_0 = -1.198$, all Petersen's models with 
$Z = 0.02$ fall in the range $\Pi_1 / \Pi_0 = 0.771 - 0.773$, quite different from our
measured value of 0.759. We suspect that this difference is due to the fact that Petersen's 
models were calculated in the absence of rotation, while Altair is an extremely rapid rotator. 
We can estimate the frequency shift due to rotation by using Saio's (1981) polytropic models
(see also Perez Hernandez et al. 1995 for a similar application). We take the frequency shift
to be 
\begin{equation}
\nu_{n00} - \nu_{n0}^{(0)} = ( Z_{n0} + X_0 ) \frac{\nu_{rot}^2}{\nu_{n0}^{(0)}}
\end{equation}
where $\nu_{nlm}$ is the frequency of the mode with radial order $n$, degree $l$, and azimuthal
order $m$, $\nu_{nl}^{(0)}$ is the frequency of the same mode in a non-rotating star, 
$\nu_{rot}$ is the stellar rotational frequency (rigid body rotation is assumed), and the 
coefficients $X_0$ and $Z_{nl}$ are structural coefficients.
Here $X_0 \approx 1.333$ and the the $Z_{n0}$ are estimated by interpolation from the
values in Saio (1981). The result indicates frequency shifts in the direction of lower frequencies,
on the order of a few percent for
the fundamental mode and nearly 10\% for the first harmonic. In addition, the prescription of
Saio does not take into account departures from rigid body rotation that likely occur for
extremely high rotation rates (in this case, for example, $\Pi_0 / \Omega \approx 0.15$). It
is thus unsurprising that the period ratios from Petersen do not correspond well to our measured
values; perhaps the larger surprise is that the period of the fundamental mode seems to fit so well!
We note without comment that assuming $M = 1.8 M_\sun$ and fitting to Petersen's evolutionary 
tracks in the $(\log \Pi_0, \log \rm Age)$ domain yields an age 
for Altair of $630 \pm 50$ Myr, which in turn is in accord with the A7~IV-V spectral
type generally assigned to Altair.

\section{Conclusions}

We have determined that the bright A7~IV-V star Altair is an oscillating variable of the
$\delta$ Scuti type, and that its three most significant oscillation frequencies correspond 
well to the fundamental radial mode and its first two overtones. Four additional frequencies
are present at lower amplitudes, and these may represent nonradial modes. Our result
suggests that oscillation may in fact be present in many putatively ``non-oscillating''
stars in the instability strip.

On the basis of the observed frequencies and the model fits, several specific points are clear: 
\begin{enumerate}
\item As suggested by the theoretical P-L relation of Breger \& Bregman (1975), 
the highest-amplitude mode is the radial fundamental mode. The amplitude of this
oscillation, 0.5~ppt, is several times smaller than is typically detected in
non-HADS $\delta$ Scuti stars from ground-based observations. The first and second
overtones are also apparent.
\item It has been suggested that (Breger 1982) that slow rotation is a necessary (but
not sufficient) condition for the development of large-amplitude oscillations in 
$\delta$ Scuti stars. Our result supports this contention.
\item A number of authors (see, {\it e.g.} Rasmussen et al. 2002 and references therein)
have tried to understand the excitation mechanism in $\delta$~Scuti stars by searching
for systematic differences between variable and non-variable stars. Our result suggests
that such searches may be complicated by the fact that ``non-variable'' stars may in
fact be variable at levels undetectable from the ground. In fact, the division between
``variable'' and ``non-variable'' stars in the instability strip may truly be an 
observational (detection technology-limited) one, with the objects in
question occupying a broad continuum of photometric amplitudes.
\item Non-rotating models should not be expected to yield accurate representations of 
oscillation frequencies or ratios in rapidly rotating $\delta$ Scuti stars. There is 
a significant need for calculations of oscillation frequencies that incorporate the
effects of rotation.

\end{enumerate}

\acknowledgments

We are grateful to all the people whose efforts contributed to the successful use
of the WIRE satellite. DLB, HB, HLP, and JM acknowledge support from NASA (NAG5-9318) and
from the US Air Force Academy. AR and TRB are supported by the Australian Research
Council and HK by the Danish National Research Foundation through its establishment
of the Theoretical Astrophysics Center.




\clearpage
\end{document}